\documentclass[11pt,twoside]{article}
\usepackage{asp2014}

\aspSuppressVolSlug
\resetcounters

\begin{document}
\title{Planetary Bistatic Radar}
\author{M.~Brozovi{\'c},$^1$ B.~J.~Butler,$^2$ Jean-Luc~Margot,$^3$ 
        Shantanu~P.~Naidu,$^4$ and  T.~Joseph~W.~Lazio$^5$
\affil{$^1$Jet Propulsion Laboratory, California Institute of
              Technology, Pasadena, \hbox{CA}, \hbox{USA};
              \email{Marina.Brozovic@jpl.caltech.edu}}
\affil{$^2$National Radio Astronomy Observatory, Socorro, \hbox{NM};
              \hbox{USA}; \email{bbutler@nrao.edu}}
\affil{$^3$University of California, Los Angeles, Los Angeles,
              \hbox{CA}, \hbox{USA}; \email{jlm@epss.ucla.edu}}
\affil{$^4$Jet Propulsion Laboratory, California Institute of
              Technology, Pasadena, \hbox{CA}, \hbox{USA}; \email{Shantanu.P.Naidu@jpl.caltech.edu}}
\affil{$^5$Jet Propulsion Laboratory, California Institute of
              Technology, Pasadena, \hbox{CA}, \hbox{USA}; \email{Joseph.Lazio@jpl.caltech.edu}}}

\paperauthor{Marina Brozovi{\'c}}{Marina.Brozovic@jpl.caltech.edu}{}%
            {Jet Propulsion Laboratory, California Institute of Technology}%
            {}%
            {Pasadena}{CA}{91109}{USA}
\paperauthor{Bryan J.~Butler}{bbutler@nrao.edu}{}%
            {National Radio Astronomy Observatory}%
            {}%
            {Socorro}{NM}{}{USA}
\paperauthor{Jean-Luc Margot}{jlm@epss.ucla.edu}{}%
            {University of California, Los Angeles}%
            {Dept.\ of Earth, Planetary, and Space Sciences}%
            {Los Angeles}{CA}{90095}{USA}
\paperauthor{Shantanu~P.~Naidu}{Shantanu.P.Naidu@jpl.caltech.edu}{}%
            {Jet Propulsion Laboratory, California Institute of Technology}%
            {}%
            {Pasadena}{CA}{91109}{USA}
\paperauthor{T.~Joseph~W.~Lazio}{Joseph.Lazio@jpl.caltech.edu}{}%
            {Jet Propulsion Laboratory, California Institute of Technology}%
            {Interplanetary Network Directorate}%
            {Pasadena}{CA}{91109}{USA}

\begin{abstract}
Planetary radar observations offer the potential for probing the
properties of characteristics of solid bodies throughout the inner
solar system and at least as far as the orbit of Saturn.  In addition
to the direct scientific value, precise orbital determinations can be
obtained from planetary radar observations, which are in turn valuable
for mission planning or spacecraft navigation and planetary defense.
The next-generation Very Large Array would not have to be equipped
with a transmitter to be an important asset in the world's planetary
radar infrastructure.  Bistatic radar, in which one antenna transmits
(e.g., Arecibo or Goldstone) and another receives, are used commonly
today, with the Green Bank Telescope (GBT) serving as a receiver.  The
improved sensitivity of the ngVLA relative to the GBT would improve
the signal-to-noise ratios on many targets and increase the accessible
volume specifically for asteroids.  Goldstone-ngVLA bistatic
observations would have the potential of rivaling the sensitivity of
Arecibo, but with much wider sky access.
\end{abstract}

\section{Introduction}\label{sec:radar.intro}

Planetary radar observations have been used to probe the surfaces of
all of the planets with solid surfaces and many smaller bodies in the
solar system \citep{o93,o03}, delivering information on their
spins, orbital states, and surface and subsurface electrical
properties and textures.  Notable findings include characterizing the
distribution of water at the south pole of the Moon
\citep{1997Sci...276.1527S,2003Natur.426..137C}, the first indications
of water ice in the permanently shadowed regions at the poles of Mercury
\citep{1992Sci...258..635S,1994Natur.369..213H}, polar ice and
anomalous surface features on Mars \citep{1991Sci...253.1508M},
establishing the icy nature of the Jovian satellites
\citep{1978Icar...34..268O}, and the initial characterizations of
Titan's surface \citep{1990Sci...248..975M,2003Sci...302..431C}.  In multiple cases, the
ground-based radar observations have served as the foundation for a
subsequent space-based mission.

Radar observations are currently conducted in the S~band ($\approx
2.3$~GHz, Arecibo Observatory) and~X~band ($\approx 8.5$~GHz, the Deep
Space Network's Goldstone Solar System Radar [GSSR]), and future radar observations
may also be conducted in the Ka~band ($\sim 30$~GHz).  All of the
planetary radar bands could be within the frequency coverage of the
next-generation Very Large Array ({ngVLA}).  As we discuss in more detail below
(\S\ref{sec:radar.sensitivity}), the ngVLA need not be equipped with a
transmitter to provide a powerful enhancement to planetary radar
capabilities.  Indeed, many of the results summarized previously
involved \emph{bistatic} observations in which the radar transmissions
originated from one  antenna and were received by a separate antenna.

\cite{bcdg04} previously considered the use of the Square Kilometre
Array (SKA) as a receiver for a bistatic system.  Since the time of
their paper, there have been a number of developments, including
multiple radar instruments on Mars orbiters, the \textit{Cassini}
radar instrument's observations of Titan, and the MESSENGER studies
that confirmed earlier radar indications of polar ice at Mercury.
This consideration of the ngVLA capabilities is similar to the earlier
SKA consideration, but takes many of these subsequent spacecraft-based
radar results into account.

We begin by motivating the scientific measurements that could be
obtained from various target bodies by bistatic radar
(\S\ref{sec:radar.targets}), then turn to the specific benefit of
the ngVLA in the context of the radar equation
(\S\ref{sec:radar.sensitivity}), and conclude with a discussion on
radar imaging (\S\ref{sec:radar.imaging}).

\section{Target Bodies}\label{sec:radar.targets}

In this section, we review target bodies and the science motivations
for which future bistatic planetary radar observations could be relevant.

\subsection{Venus}\label{sec:radar.venus}

Venus is Earth's closest analog in the solar system in terms of its
bulk properties, yet Venus and Earth have clearly had different
evolutionary paths.  There are potentially billions of Venus analogs
in the Galaxy, and characterizing and understanding the differences
between Venus and Earth has been given additional impetus for
understanding the habitability of terrestrial-mass planets.  Venus
remains enigmatic on a variety of fundamental levels: The size of its
core is unknown; whether the core is solid or liquid is uncertain; its
atmospheric superrotation, 60$\times$ faster than the solid body, is
not understood; and the atmosphere exhibits distinctive
planetary-scale features that are stationary with respect to the solid
body.  High-precision measurements of the spin state of Venus with
radar have the potential of providing key advances in all of these
areas.  First, a measurement of the spin precession rate ($\approx
2^{\prime\prime}\,\mathrm{yr}^{-1}$) will yield a direct measurement
of the polar moment of inertia, which is unknown.  The moment of
inertia provides an integral constraint on the distribution of mass in
a planetary interior.  Apart from bulk density, it is arguably the
most important quantity needed to determine reliable models of the
interior structure of Venus, including the size of its core.  Second,
a time history of length of day (LOD) variations at the 10~ppm level
will identify the geophysical forcings responsible for spin
variations, which are primarily related to transfer of angular
momentum between the atmosphere and the solid planet.  They will
provide a crucial input to general circulation models and the key to
elucidate poorly understood phenomena such as superrotation and
stationary planetary-scale structures in the atmosphere.

Planetary radar provides a powerful tool for
monitoring planetary spin states via observations of the ``speckle
displacement effect'' or \emph{radar speckle tracking} \citep{mpjsh07}.
Radar echoes from solid bodies exhibit
spatial irregularities in the wavefront caused by the constructive and
destructive interference of waves scattered by the irregular surface.
The corrugations in the wavefront, i.e., speckles, are tied to the
rotation of the target body.  When the trajectory of the wavefront
corrugations is parallel to a roughly east-west antenna baseline,
echoes received at two receiving stations display a high degree of
correlation. The time of day and value of the time delay at the
correlation peak are directly related to the orientation and magnitude
of the spin vector of the body.  For typical solar system
observations, the speckle size ($\sim R\lambda/D$, for a target at
range~$R$, observing wavelength~$\lambda$, and diameter~$D$) is on the
order of~1~km and the high-correlation condition lasts for
approximately 30~s.

The current approach using Goldstone and the Green Bank Telescope (GBT) yields
instantaneous spin rate measurements at the 10~ppm level with X-band
transmission from the GSSR and reception at Goldstone and the \hbox{GBT}.  For example, with observations obtained between~2002
and~2012, the orientation of Mercury's spin axis has been measured
with~$5^{\prime\prime}$ precision, and measurements of the amplitude
of longitude librations have revealed Mercury has a molten core
\citep{mpjsh07,mps+12}.  The accuracy of these measurements has been
validated at the 1\% level by independent measurements obtained by the
MESSENGER spacecraft during a four-year duration \citep{mhmpp18}.  On-going observations include Venus, Europa, and Ganymede.

With a single pair of antennas, it is possible to obtain one
measurement per day when the stringent geometry and signal-to-noise
(S/N) requirements are satisfied.  However, measurements accumulate at
a slow rate because each measurement requires simultaneous scheduling
on two large radio antennas, successful transmission during the
appropriate 30-second window, and successful reception at both
antennas during the relevant 30-second windows.  In order to fully
constrain the spin axis orientation, it is imperative to secure
observations at a variety of baseline orientations, which typically
takes several years.  An instrument such as the ngVLA would open up
the possibility of securing up to 168 independent measurements in a
single 20-minute session with the antennas located in the plains of
San Agustin.  At conjunction, the individual antenna S/N ratio and the
correlation S/N ratio would exceed 100.  Although the range of
baseline orientations between array elements and Goldstone would
remain small at any given observation epoch, the number of independent
estimates of the correlation properties would improve the quality of
the spin state determination by a factor of $\sqrt{N}$.  Because the
measurements are instantaneous, LOD variations that occur on 30-minute
timescales would be detectable, which would place strong constraints
on the mechanisms of angular momentum transfer.

Consequently, the ngVLA would enable (1)~improved determination of the
spin axis precession and therefore moment of inertia and core size;
and (2)~improved quantification of the amplitude of LOD variations on daily, seasonal, and secular timescales, providing strong constraints on the dynamics of the atmosphere and its interactions with the solid planet, and exploring a regime that may be common on exoplanets.  

\subsection{Asteroids}\label{sec:radar.asteroids}

\begin{figure}[tbh]
  \centering
  \includegraphics[width=0.95\textwidth]{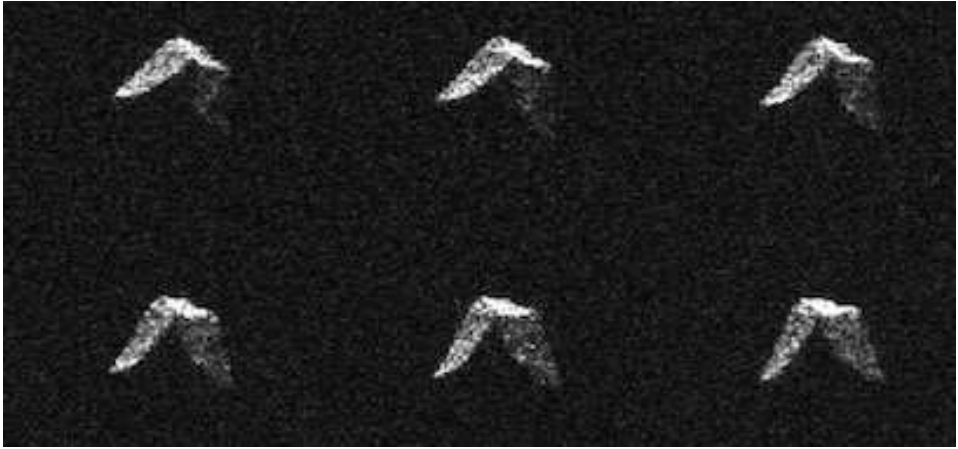}  
  \vspace*{-1ex}
  \caption{Radar observations of the near-Earth asteroid 2017~BQ6.
    Its rotation is apparent, as are the sharp, angular sides.  A
    bright spot particularly apparent in the lower, middle panel may
    be a few meter-scale boulder on the surface.  The sharpness of
    this asteroid's structure is currently unexplained.
    The time delay
    (range) increases from top to bottom, and Doppler frequency
    increases from left to right.  (I.e., the top of the figure is
    closest to Earth, and the image can be considered to be a
    ``top-down'' view.)  The color scale shows the echo power strength in units
    of standard deviation.}
  \label{fig:radar.2017bq6}
\end{figure}

Radar observations of asteroids provides information on their
sizes, shapes, spin states, surface properties, masses, bulk
densities, orbits, and the presence of satellites (Figure~\ref{fig:radar.2017bq6}).
Recent improvements in transmitter capabilities have resulted in
obtaining meter-scale spatial resolutions on various near-Earth
asteroids, resolutions comparable to those obtained by spacecraft
(either for fly-bys or orbiting).  Thus, radar observations 
complement the fewer, but often more comprehensive spacecraft measurements.

There have been a series of comprehensive reviews on radar
observations of both near-Earth and Main Belt asteroids 
\citep{mor+99,ohb+02,o03,bcdg04,bbgtm15,nbmbt16}.  We do not repeat
that material here, but focus on specific aspects relative to the
\hbox{ngVLA}.  

The motivation for radar observations of asteroids is three-fold.
First, asteroids represent primitive remnants of the early solar
system, and their properties and orbits provide constraints on the
formation and evolution of the solar system.  Second, they represent
targets for spacecraft \citep[e.g.,][]{2014Icar..235....5C}, for which
orbital information and the presence of satellites are essential for
mission planning and, for sample return missions, characterization of
the surfaces is valuable.  Finally, precise knowledge of their orbits
is essential to assess the extent to which they might represent impact
hazards to the Earth \citep{NEO_hazard}, a topic that has increased in
visibility over the past decade, and for which a ``National Near-Earth
Object Preparedness Strategy and Action Plan'' has been issued
\citep{DAMIEN2018}.  In particular, the orbits determined from radar
observations are sufficiently precise that they can be used to assess
whether a near-Earth asteroid presents any risk of colliding with the Earth over the
next several decades to a century \citep[e.g.,][]{og04}.

Specifically for near-Earth asteroids, bistatic radar observations can
be valuable in two respects.  First, for objects with close approaches
to the Earth (short round-trip light travel times), it can be
difficult or impossible to switch a radar facility from transmitting
to receiving rapidly enough.  Bistatic radar observations either
simplify the observations or enable them for objects on extremely
close approaches.  Second, the increased sensitivity of the ngVLA
would increase the range to which near-Earth asteroids could be
targeted for radar observations, particularly for targets that are
outside of the declination range of the Arecibo Observatory.
Particularly from the perspective of planetary defense, obtaining
orbits for as many near-Earth asteroids, and especially those
classified as ``potentially hazardous'' is valuable, and the larger
the volume that is accessible, the more asteroids that can be
targeted.  We return to this topic, in quantitative detail, in Section~\ref{sec:radar.sensitivity}.

As quantitative estimates, we consider the improvement in range that
the ngVLA might offer over the Green Bank Telescope, which is also
used as the receiving element for bistatic radar.  (See also
\citealt*{nbmbt16}.)  If a subset of the ngVLA can be used for
bistatic radar reception such that a sensitivity of \textbf{three}
times that of the current \hbox{GBT} is obtained, it would more than
\textbf{double} the accessible volume (increase the range by a factor
of~30\%) for near-Earth asteroid observations; if the sensitivity is
\textbf{five} times that of the current \hbox{GBT}, it would more than
\textbf{triple} the accessible volume (increase the range by a factor
of~50\%).  Not only could Goldstone-ngVLA bistatic observations rival
those of Arecibo, they would provide access to a much larger fraction
of the sky.

Beyond the simple improvement in sensitivity offered by the ngVLA
(\S\ref{sec:radar.sensitivity}), its antenna distribution offers the
promise of improved shape modeling and spin state determinations via
radar speckle tracking  \citep{bkb+10}.  
Speckle tracking of asteroids operates in a fundamentally different
regime than the case of Mercury or Venus \citep{mpjsh07}.  The
general inability to predict the speckle trajectory in the asteroid
case requires an observing configuration in which the speckle size is
larger than the antenna baseline, otherwise speckles observed at
different stations would not correlate.  As a result, the ratio of
speckle size to antenna baseline, which determines the fractional
precision of the estimates, is three orders of magnitude larger for
asteroids than it is for Mercury or Venus.  
The
VLA has a dense set of antennas, but few asteroids approach the Earth
sufficiently close that the VLA can be used.  For example, for a 100~m
object, it must be within about 10\% of the lunar distance for the
resulting speckle pattern to be comparable in scale to the
\hbox{VLA}.  Conversely, the VLBA has much longer baselines, allowing
use of the technique to larger distances, but it has few antennas, so
the number of speckle measurements that could be made is few.  With a
relatively dense network of antennas and antenna separations to of
order 100~km, the ngVLA could be used for objects approaching to
within one lunar distance ($10^{-3}$~au), for which the number of
objects is higher.

Finally, the dynamics of the Sun-Earth-Moon system allow for small
asteroids to be captured into meta-stable geocentric orbits.  Various
predictions are that there should be a population of meter-scale
``temporarily-captured orbiters'' or ``mini-moons,'' and at least one
such mini-moon, 2006~RH120, has been detected
\citep{kkp+09,bjg+14,jbb+18}.  The advent of future large
scale-surveys, such as the Large Synoptic Survey Telescope (LSST), may
result in several more being found.  Due to their small size, such
observations are extremely challenging, but feasible, as 2006~RH120
has been detected from Goldstone \citep{ bbbgsl16}.  Not only do
mini-moons have small radar cross sections, they are relatively close
($\approx 3$~s round-trip light travel time).  As noted above, with
such short light travel times, bistatic observations would be required
to avoid subjecting the transmitter to frequent power fluctuations.

\subsection{Icy Satellites/Ocean Worlds}\label{sec:radar.oceanworlds}

Ground-based radar observations provided some of the first clear
evidence for the icy surfaces of the Galilean satellites and
subsequent characterization \citep[e.g.,][]{1977Sci...196..650C,
  1978Icar...34..268O,1992JGR....9718227O,2001Icar..151..167B},
demonstrating the capability to probe several meters into the
surface.  There have been radar observations of Saturnian satellites
as well, though these are even more challenging due to Saturn's
greater distance \citep{1990Sci...248..975M,2003Sci...302..431C,2007Icar..191..702B}.
Subsequent
spacecraft investigations have provided clear evidence that at least
some of the moons of Jupiter (Europa, Ganymede, Callisto) and Saturn
(Enceladus, Titan), and potentially moons of Uranus (Ariel) harbor
sub-surface oceans \citep{2016JGRE..121.1378N}.  As a consequence, the planned Europa Clipper mission would carry a
radar designed to probe and constrain the thickness of the icy shell
of Europa.

However, spacecraft missions are infrequent, plausibly only two
missions might fly to the outer solar system in a decade, and a radar
instrument might not be part of the spacecraft's payload.  By
contrast, ground-based radar can potentially happen essentially
annually, near an outer planet's opposition, when the distance to
the icy satellite is minimized.

The orbits of the Galilean satellites of Jupiter are affected by tidal
interactions with Jupiter, due to their relatively small semi-major
axes.  Of particular interest are the tidal responses of Io and Europa
as the tidal response of Io is related to the heat dissipation
responsible for its active volcanism and the tidal response of Europa
is related to the depth of its sub-surface ocean Tidal dissipation is
parameterized by $k_2/Q$, where $k_2$ is the Love number, which is a
measure of the amplitude of tidal response in the body, and~$Q$ is the
quality factor or a measure of the viscous damping in the body.  Both
quantities are related to the properties of Jupiter's or satellites'
interiors, and the ratio~$k_2/Q$ quantifies how the bulge raised by
the satellite, or on the satellite, leads or trails its ``precursor.''
The Juno mission will provide estimates of $k_2$, which can be
combined with the radar ranging, to estimate the ratio~$k_2/Q$.
Moreover, radar ranging measurements have the advantage of being able
to be carried out indefinitely, while the Juno mission is of a limited
duration.  (At the time of writing, the Juno prime science mission
terminates in~2021 June.)

\cite{lakh09} used astrometric data to suggest that orbits of Io, Europa, and Ganymede have shifted due to tidal acceleration by~55~km, $-125$~km, and~$-365$~km, respectively, over a period of~116~years. The highest precision astrometric measurements have a resolution of~75~km and originate from mutual occultations and eclipses.

\begin{figure}[bth]
  \centering
  \includegraphics[width=0.95\textwidth]{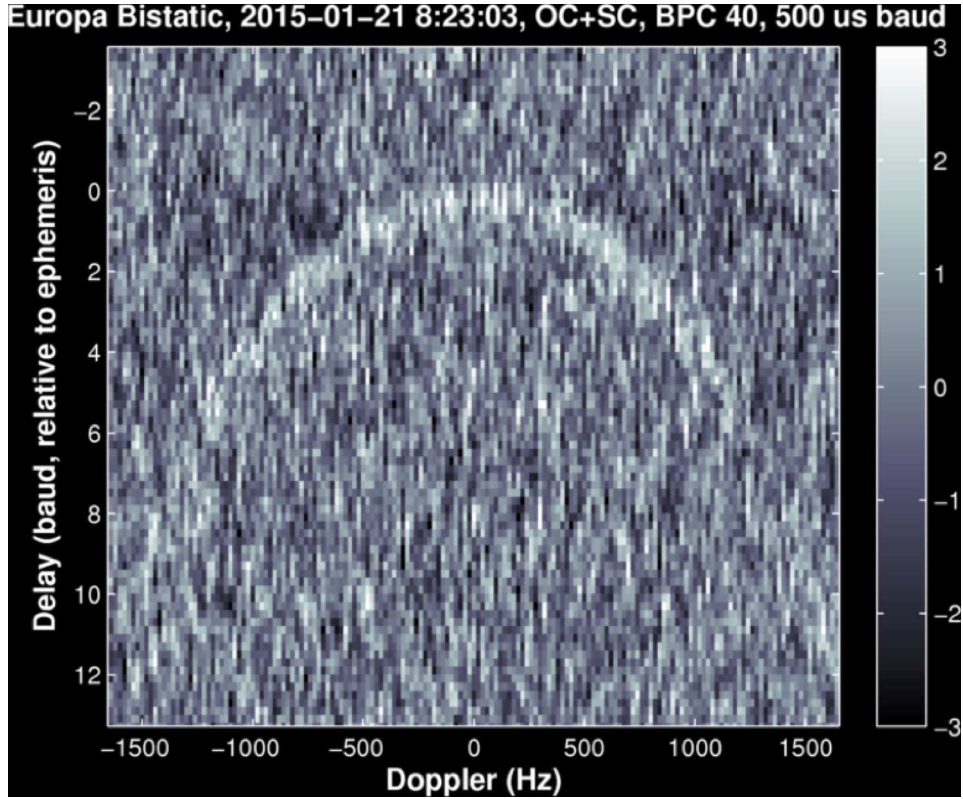}
  \vspace*{-1ex}
  \caption{Goldstone Solar System Radar-Green Bank Telescope
    delay-Doppler image of Europa. The time delay
    (range) increases from top to bottom, and Doppler frequency
    increases from left to right.  (I.e., the top of the figure is
    closest to Earth, and the image can be considered to be a
    ``top-down'' view.)  The range resolution is 500~$\mu$s
    or~75~km. The color scale shows the echo power strength in units
    of standard deviation. The scale has been saturated at~3 units in
    order to enhance the echo outline.  With its higher sensitivity, the ngVLA would offer higher
    signal-to-noise ratios and higher ranging precision.}
  \label{fig:radar.europa}
\end{figure}

The Arecibo planetary radar can measure a line-of-sight distance (or
range) to Io with~10~km precision and distances to Europa, Ganymede,
and Callisto with~1.5~km precision. However, Jupiter is only
observable from Arecibo six out of every 12~years because of the
constraints of the Arecibo's antenna pointing (declination range
$-1^\circ$ to~$+38^\circ$).
In~2015, GSSR-GBT bistatic radar demonstrated the capability to obtain
ranging measurements of the Galilean moons. Both antennas are fully
steerable and allow observations on a yearly basis. A range to Europa
was measured with~75~km precision (Figure~\ref{fig:radar.europa}), comparable to the
highest precision optical astrometry.  However, ranging to Io was not
possible with in this bistatic configuration due to low echo strength.

If ngVLA achieves 3 times the sensitivity of the \hbox{GBT}, it would
likely be able to achieve yearly ranging measurements of Europa with
sub-10~km precision. Furthermore, it would possible to measure a
distance to Io with~100~km precision. If a five-fold sensitivity of
the ngVLA materializes, ranging precisions of~15~km--30~km for Io
could be obtained.  These measurements will contribute to the
maintenance of highly accurate ephemerides of the Galilean satellites
that would lead to continued improvements in the constraints on tidal
dissipation ($k_2/Q$) and that could enhance, and potentially enable,
future missions to these bodies (e.g., monitor Io's volcanism, explore
Europa's ice shell for biosignatures).

\subsection{Comets and Interstellar Objects}\label{sec:radar.comets}

Much like the case for asteroids, radar observations of comets can
provide information on the size, shape, and spin state of comet
nuclei.

For instance, the Arecibo radar observed comet 103P/Hartley~2 shortly
before NASA's EPOXI spacecraft encountered it in~2010 November.  The
radar observations determined that the comet nucleus has a bi-lobed
shape, a result confirmed by the spacecraft \citep{hnhgt11}.
Also, much like the case for asteroids, the radar
observations of cometary nuclei complement the fewer, but more
incisive spacecraft measurements---approximately five times as many
comets have been detected by radar observations as have been visited
by spacecraft.

The recent recognition of the first interstellar object, 1I/2017 U1
`Oumuamua, suggested that such objects might have extremely low
optical albedos and at least this first object appeared to have a
large aspect ratio, potentially in excess of 5:1 \citep{mwm+17}.
The number of identified interstellar objects may increase in the
future as additional wide-field surveys occur, particularly if the
survey strategies explicitly account for the potential trajectories of
interstellar objects.  Notably, \cite{jlr+17} predict that there may
be as many as $10^4$ such objects within the orbit of Neptune at any
given time.  If an interstellar object did have a trajectory that took
it sufficiently close to Earth to warrant radar observations,
constraints on its properties would be invaluable.

\section{The ngVLA and the Radar Equation}\label{sec:radar.sensitivity}

The increased sensitivity of the ngVLA would expand the set of targets
for traditional bistatic delay-Doppler planetary radar.  The classic
radar equation is that the received power~$P_R$ is \citep[e.g.,][]{bho+99}
\begin{equation}
P_R = \frac{P_T G_T G_R \lambda^2 \sigma}{(4\pi)^3 R^4},
\label{eqn:radar}
\end{equation}
where $P_T$ is the power of the transmitter; the gains of the
transmitting and receiving antennas are $G_T$ and~$G_R$, respectively;
$\lambda$ is the operational wavelength; $\sigma$ is a measure of the
radar cross section of the target body; and the range (distance) to
the target is $R$.  More concisely, the signal-to-noise ratio of
radar observations scales as $R^{-4}$.  This $R^{-4}$ dependence can
be understood as the product of two inverse square laws.  The
transmissions from the transmitting antenna to the target body suffer a
$R^{-2}$ loss by the inverse square law.  By Huygen's principle, the
target body re-radiates, and these emissions suffer an additional
$R^{-2}$ loss by the inverse square law.

With a fixed radar transmitter power (and antenna gain), the
signal-to-noise ratio can only be improved by using increasing the
gain (i.e., sensitivity) of the (bistatic) receiving element.  For the
\hbox{ngVLA} to participate in this kind of radar observation, it
would have to have a ``phased array'' mode, in which voltages from
the individual antennas are summed, after applying the appropriate
time delays, so that the array appears as an effective single
aperture.  A phased array mode would also be valuable for observations
of pulsars and for very long baseline interferometry (VLBI) imaging.

\section{Radar Imaging}\label{sec:radar.imaging}

Existing planetary radar observations with the VLA have been used to
image the radar return and make plane-of-the-sky astrometry
measurements.  Such \emph{bistatic radar aperture synthesis} has produced
spectacular images that convincingly demonstrated the presence of
water ice at the poles of Mercury and Mars.  With a factor of
\textbf{five} better angular resolution, the ngVLA could produce a
correspondingly better linear resolution on the surface of target
bodies.  Alternately, an improved angular resolution opens the
possibility of producing resolved images of smaller objects or higher
spatial resolution bistatic radar aperture synthesis on planets.

For
example, \cite{1994Icar..111..489D} imaged 4179 Toutatis with the VLA
in its A configuration.  The asteroid was at a distance of~0.063~au,
and the VLA's angular resolution corresponded to a linear resolution
of approximately 10~km on the asteroid.  They found that the asteroid
showed clearly distinct residual radar features, suggestive of a
bi-lobed structure, but their estimate of the separation of these
features was limited by the VLA's beam size.  With the improved
angular resolution of the \hbox{ngVLA}, features with scales of
about~2~km would have been distinguishable.

\bigskip
{\small%
We thank L.~Benner, M.~Busch, and P.~Taylor for helpful comments.
This work made use of NASA's Astrophysics Data System Abstract
Service.  
Part of this research was carried out at the Jet Propulsion
Laboratory, California Institute of Technology, under a contract with
the National Aeronautics and Space Administration.
The National Radio Astronomy Observatory is a facility of the
National Science Foundation operated under cooperative agreement by
Associated Universities, Inc.}

\end{document}